\documentclass[aps,prd,print,groupedaddress,nofootinbib,eqsecnum,onecolumn]{revtex4}
\usepackage{eurosym}
\usepackage{amsmath}
\usepackage{amsthm}
\usepackage{amssymb}
\usepackage{slashed}
\usepackage{hyperref}
\usepackage{epstopdf}
\usepackage[english]{babel}
\usepackage[latin1]{inputenc}
\usepackage{amsmath, amsthm, amssymb}
\usepackage[pdftex]{graphicx}
\usepackage{amssymb}
\usepackage{latexsym}
\usepackage{amstext}
\usepackage{epstopdf}
\usepackage{floatflt}
\usepackage{hyperref}
\usepackage{enumitem}
\usepackage{multirow}
\usepackage{pdflscape}
\usepackage{makecell}
\usepackage{nopageno}

\numberwithin{equation}{section}
\begin{document}

\begin{titlepage}
\begin{center}

\vskip 3.0cm

{\bf \huge Jordan meets Freudenthal}\\
\vskip 1.0cm
{\bf \Large A Black Hole Exceptional Story}

\vskip 3.0cm

{\bf \large Alessio Marrani}

\vskip 40pt

{\it Instituto de F\'{\i}sica Teorica, Dep.to de F\'{\i}sica,\\
Universidad de Murcia, Campus de Espinardo, E-30100, Spain}\\
\texttt{alessio.marrani@um.es},

\vskip 5pt

\vskip 40pt


\vskip 40pt


\vskip 30pt



\end{center}

\vskip 95pt

\begin{center} {\bf ABSTRACT}\\[3ex]\end{center}

Within the extremal black hole attractors arising in ungauged $\mathcal{N}%
\geqslant 2$-extended Maxwell Einstein supergravity theories in $3+1$
space-time dimensions, we provide an overview of the stratification of the
electric-magnetic charge representation space into \textquotedblleft
large\textquotedblright\ orbits and related \textquotedblleft moduli
spaces\textquotedblright , under the action of the (continuous limit of the)
non-compact $U$-duality Lie group. While each \textquotedblleft
large\textquotedblright\ orbit of the $U$-duality supports a class of
attractors, the corresponding \textquotedblleft moduli
space\textquotedblright\ is the proper subspace of the scalar manifold
spanned by those scalar fields on which the Attractor Mechanism is inactive.
We present the case study concerning $\mathcal{N}=2$ supergravity theories
with symmetric vector multiplets' scalar manifold, which in all cases (with
the exception of the minimally coupled models) have the electric-magnetic
charge representation of $U$-duality fitting into a reduced Freudenthal
triple system over a cubic (simple or semi-simple) Jordan algebra.

\vskip 65pt

\begin{center}
Plenary talk presented at the \textit{34th International Colloquium on Group Theoretical Methods in Physics},\\Strasbourg, July 18-22, 2022
\end{center}





%
\vfill

\end{titlepage}

\newpage \setcounter{page}{1} \numberwithin{equation}{section}

\section{Introduction\label{Intro}\protect\smallskip}

Within the theory of dynamical (dissipative) systems, an \textit{attractor}
is defined by a \textit{fixed point} of the evolution flow of the system
itself, describing the equilibrium state and its stability features. In
general, when approaching an attractor, the orbits of the dynamical
evolution lose all memory of their initial conditions, but nonetheless the
overall dynamics remains strictly deterministic.

Within Maxwell-Einstein-scalar theories endowed with local supersymmetry in $%
3+1$ space-time dimensions, attractors were firstly discovered within the
class of extremal black hole solutions preserving half of the
supersymmetries, in presence of $\mathcal{N}=2$ spinor supercharges. This
led to the discovery of the so-called \textit{Attractor Mechanism} (AM),
governing the dynamics of evolution of the scalar fields in the black hole
background \cite{AM-Ref1}-\nocite{AM-Ref2,AM-Ref3}\cite{AM-Ref4}. We will now review the basics of AM in such a
framework.

As far as propagating (i.e., dynamical) massless fields are concerned,
linearly realized $\mathcal{N}=2$ local supersymmetry in $3+1$ space-time
dimensions admits three multiplet representations (see \textit{e.g.} \cite%
{N=2-Big} for a general treatment and a list of Refs.) :

\begin{enumerate}
\item one \textit{gravity} multiplet, whose maximal helicity is $2$, given
by
\begin{equation}
\left( V_{\mu }^{a},\psi ^{A},\psi _{A},A^{0}\right) ,  \label{g-mult}
\end{equation}%
where the \textit{Vielbein} one-form $V^{a}$ (together with the
spin-connection one-form $\omega ^{ab}$) relates to the graviton ($a=0,1,2,3$%
), $\psi ^{A},\psi _{A}$ are $SU(2)$-doublets of spinor one-forms (usually
named gravitinos; $A=1,2$, with the upper and lower indices respectively
denoting right and left chirality, \textit{i.e.} $\gamma _{5}\psi
_{A}=-\gamma _{5}\psi ^{A}$), and $A^{0}$ denotes the Maxwell gauge boson
1-form potential usually named \textit{graviphoton}.

\item $n_{V}$ \textit{vector} multiplets, whose maximal helicity is $1$,
given by ($I,i=1,...,n_{V}$)
\begin{equation}
\left( A^{I},\lambda ^{iA},\overline{\lambda }_{A}^{\overline{i}%
},z^{i}\right) ,  \label{v-mult}
\end{equation}%
each containing a gauge boson one-form $A^{I}$ , a $SU(2)$-doublet of
zero-form spinors $\lambda ^{iA},\overline{\lambda }_{A}^{\overline{i}}$
(usually named gauginos), and a complex scalar field (zero-form) $z^{i}$.
The $z^{i}$'s coordinatize a complex manifold $M_{n_{V}}$, of complex
dimension $n_{V}$, which is endowed with a projective \textit{special K\"{a}%
hler} structure by supersymmetry.

\item $n_{H}$ \textit{hypermultiplets}, whose maximal helicity is $1/2$,
given by ($\alpha =1,...,2n_{H}$)
\begin{equation}
\left( \zeta _{\alpha },\zeta ^{\alpha },q^{u}\right) ,  \label{h-mult}
\end{equation}%
each containing a pair of zero-form spinors $\zeta _{\alpha },\zeta ^{\alpha
}$ (named hyperinos), and four real scalar fields $q^{u}$ ($u=1,...,4n_{H}$%
), which coordinatize a quaternionic manifold $\mathcal{Q}_{n_{H}}$ (of
quaternionic dimension $n_{H}$).
\end{enumerate}

When there is no gauging of any global isometry of $M_{n_{V}}$ and/or $%
Q_{n_{H}}$, the $n_{H}$ hypermultiplets are not involved in the AM, and they
can be completely decoupled from the attractor dynamics in the black hole
background. This is a direct consequence of the supersymmetry transformation
properties of the zero-form spinor fields : the hyperinos $\zeta _{\alpha }$%
's transformations do not depend on the graviphoton $A^{0}$ nor on $A^{I}$'s
(i.e., on the Maxwell 1-form potentials), whereas gauginos $\lambda ^{iA}$'s
ones do depend on the Maxwell potentials. More precisely, when disregarding
for simplicity's sake the fermionic and gauging terms, the supersymmetry
transformations of hyperinos read \cite{N=2-Big}
\begin{equation}
\delta \zeta _{\alpha }=i\mathcal{U}_{u}^{B\beta }\partial _{\mu
}q^{u}\gamma ^{\mu }\varepsilon ^{A}\epsilon _{AB}\mathbb{C}_{\alpha \beta },
\label{hyperinos}
\end{equation}%
implying that the values of the quaternionic scalar fields $q^{u}$ in the
asymptotical(ly flat,) spacial background are unconstrained, and thus they
can vary continuously within $Q_{n_{H}}$. In other words, the hyperscalars $%
q^{u}$'s are \textit{moduli} of the system in absence of gauging.

Consequently, in order to keep the framework as simple as possible, we can
totally disregard hypermultiplets, and this actually does not imply any loss
of generality, at least when ungauged theories are considered. Thus, we
consider asymptotically flat, sphedrically symmetric, static, dyonic
extremal black hole solutions of $\mathcal{N}=2$-extended supergravity, in
which the gravity multiplet (\ref{g-mult}) is coupled to $n_{V}$ vector
multiplets (\ref{v-mult}). Since there is no dependence of the black hole
metric on time and azimuthal and polar angles, the unique coordinate
characterizing the dynamical evolution of the $n_{V}$ complex scalar fields
(one for each vector multiplet) is the radial coordinate : the AM states
that, when approaching the event horizon of the black hole, one can always
find a solution of the scalar flow such that the scalars dynamically run
into fixed points, acquiring values which only depend on (the ratios of) the
electric and magnetic charges of the black hole (respectively denoted by $%
q_{\Lambda }$ and $p^{\Lambda }$, with $\Lambda =0,1,...,n_{V}$), which are
conserved quantities due to the overall $U(1)^{n_{V}}$ gauge symmetry of the
system itself and are arranged into the symplectic vector%
\begin{equation}
\mathcal{Q}:=(p^{\Lambda },q_{\Lambda })^{T}.\label{Q}
\end{equation}%
Such near-horizon configurations of the scalar fields are completely
independent on the boundary conditions of the corresponding dynamics, namely
on the spacial asymptotical values of the scalars. Consequently, the
dynamical system describing the scalar flow completely loses memory of its
initial data, because the dynamical evolution is \textquotedblleft
attracted\textquotedblright\ to some fixed configuration points, depending
on the electric and magnetic charges \textit{only}. Note that there are no
attractors in \textquotedblleft pure\textquotedblright\ $\mathcal{N}=2$
supergravity, since the $\mathcal{N}=2$ gravity multiplet (\ref{g-mult}) has
no scalar fields (in fact, the Reissner-Nordstr\"{o}m extremal black hole
background is scalarless).

In presence of (linearly realized) local supersymmetry, extremal black holes
can be interpreted as BPS (Bogomol'ny-Prasad-Sommerfeld)-saturated \cite{BPS}
solutions, in the low-energy, effective field theory limit of
higher-dimensional, UV-complete theories, such as $\left( 9+1\right) $%
-dimensional superstrings or $\left( 10+1\right) $-dimensional $M$-theory
\cite{GT-p-branes}. As class of solutions to the Maxwell-Einstein equations
of motion, the extremal black holes under considerations are determined by
their (asymptotical) ADM mass \cite{ADM}, by the electrical and magnetic
charges (defined by integrating the fluxes of related field strengths'
2-forms over a two-sphere at infinity), and by the asymptotical values of
the $n_{V}$ complex scalar fields. Thus, the AM implies that the extremal
black holes become \textquotedblleft bald\textquotedblright , i.e. they lose
all their \textquotedblleft scalar hair\textquotedblright\ in the
near-horizon limit; in other words, when the extremal black hole metric
approaches the conformally flat Bertotti-Robinson $AdS_{2}\otimes S^{2}$
metric \cite{BR1,BR2}, it is completely characterized only by electric and
magnetic charges, but \textit{not} by the continuously-varying asymptotical
values of the scalar fields.

A major breakthrough in the study of AM was achieved in \cite{FGK}, in which
the fixed points of the scalar dynamics in the extremal black hole
background were characterized as critical points of a suitably defined
\textquotedblleft black hole effective potential\textquotedblright\ $V_{BH}$%
, in general being a strictly positive definite function of the $2n_{V}$
real scalars $\phi ^{a}$ (corresponding to $n_{V}$ complex scalar fields)
and of the $2n_{V}$ electric and magnetic (real) charges : $%
V_{BH}=V_{BH}\left( \phi ,\mathcal{Q}\right) $ For a fixed set of e.m.
charges $\mathcal{Q}$ (\ref{Q}), the non-degenerate critical points of $%
V_{BH}$ in $M_{n_{V}}$, i.e. those points in $M_{n_{V}}$ such that
\begin{equation}
\frac{\partial V_{BH}}{\partial \phi ^{a}}=0:~\left. V_{BH}\right\vert _{%
\frac{\partial V_{BH}}{\partial \phi }=0}>0,~\forall a=1,...,2n_{V},
\label{crit-points}
\end{equation}%
completely determine the values of the scalar fields in the near-horizon
limit, which depend on the electric and magnetic charges of the black hole
only. The (semi)classical Bekenstein-Hawking entropy ($S_{BH}$) - area ($%
A_{H}$) formula \cite{BH1}-\nocite{BH2,BH3}\cite{BH4} yields the extremal black hole entropy $S_{BH}$ to
be given by ($\pi $ times) the critical value of $V_{BH}$ itself :
\begin{equation}
S_{BH}=\pi \frac{A_{H}}{4}=\pi \left. V_{BH}\right\vert _{\frac{\partial
V_{BH}}{\partial \phi }=0}.  \label{BH-entropy-area-formula}
\end{equation}%
These result reduce the study of extremal black hole attractors to the study
and classification of the various classes of critical points of $V_{BH}$
which yield a non-vanishing critical value of $V_{BH}$ itself; as we will
see below, each of these classes is in $1:1$ correspondence with a $U$-orbit
supporting an attractor, and thus to an attractor \textquotedblleft moduli
space\textquotedblright .

The fluxes (over $S_{\infty }^{2}$, which exists because of the spherical
symmetry of the black hole metric) of the Maxwell 2-form field strengths
(and of their Lagrangian duals) determine the electric-magnetic charges $%
\mathcal{Q}$ (\ref{Q}) of the black hole itself, which are $2\left(
n_{V}+1\right) $ conserved quantities, where $n_{V}$ is the number of vector
multiplets. The \textquotedblleft $+1$\textquotedblright\ corresponds to the
contribution of the graviphoton Maxwell field. In the limit of real values
(which is customarily taken within supergravity, thus disregarding charge
quantization, and in particular the Dirac-Schwinger-Zwanzinger quantizations
condition for dyons), the $2\left( n_{V}+1\right) $ e.m. charges
coordinatize a vector space which is the representation space $\mathcal{Q}%
\equiv \mathbf{R}_{\mathcal{Q}}$ of the $U$-duality Lie group $G$. Onto $%
\mathbf{R}_{\mathcal{Q}}$, $G$ acts as a (maximal, non-symmetric) subgroup
of $Sp\left( 2(n_{V}+1),\mathbb{R}\right) $, the split form of the Lie group
whose Lie algebra is $\mathfrak{c}_{n_{V}+1}$ :
\begin{equation}
G\overset{\mathbf{R}_{\mathcal{Q}}}{\subsetneq }Sp\left( 2(n_{V}+1),\mathbb{R%
}\right) .  \label{emb}
\end{equation}%
Equivalently, one can state that the embedding (\ref{emb}), whose
relevancein field theory was firstly studied by Gaillard and Zumino \cite{GZ}%
, is a consequence of the fact that the (not necessarily irreducible) $G$%
-representation $\mathbf{R}_{\mathcal{Q}}$ is anti-self-conjugated (i.e.,
symplectic), by applying a general theorem of Dynkin \cite{Dynkin}.
Moreover, it should be pointed out that what we are naming as $U$-duality
Lie group $G\equiv G_{\mathbb{R}}$ is actually the (unquantized,) continuous
version of the actual $U$-duality, stringy group $G_{\mathbb{Z}}$ \cite{HT}.
This is consistent with the aforementioned (semi-)classical limit of real
charges, also taken into account by the fact that we consider $Sp\left(
2(n_{V}+1),\mathbb{R}\right) $, and not $Sp\left( 2(n_{V}+1),\mathbb{Z}%
\right) $.\medskip

Since the action of $G$ onto $\mathbf{R}_{\mathcal{Q}}$ is in general
\textit{non-transitive}, the linear representation vector space $\mathbf{R}_{%
\mathcal{Q}}$ gets stratified into disjoint classes of orbits under the
action of $G$ itself \cite{FG1,DFL,FM} : in general, a $G$-orbit $\mathcal{O}
$ is a (usually non-symmetric) homogeneous space of $G$,%
\begin{equation}
\mathcal{O}\simeq \frac{G}{\mathcal{H}}\subsetneq \mathbf{R}_{\mathcal{Q}},
\label{orb}
\end{equation}%
where the isotropy Lie group $\mathcal{H}$ is a (generally non-maximal nor
compact) subgroup of $G$ itself, and it is named \textit{stabilizer} of $%
\mathcal{O}$.

A remarkable fact, stemming from the classical invariant theory applied to
the mathematical structure of Maxwell-Einstein-scalar theories, is the
following : in all\footnote{%
The same holds for $\mathcal{N}\geqslant 3$-extended supergravity theories,
which however we will not treat here.} $\mathcal{N}=2$ supergravity theories
with homogeneous \textit{symmetric} (vector multiplets') scalar manifolds in
$3+1$ space-time dimensions, the pair $\left( G,\mathbf{R}_{\mathcal{Q}%
}\right) $ is (a suitable real form of) a $\theta $-group \textit{\`{a} la
Vinberg }\cite{Vinberg}, namely the number of nilpotent $G$-orbits in $%
\mathbf{R}_{\mathcal{Q}}$ is finite, and the ring of $G$-invariant
polynomials on $\mathbf{R}_{\mathcal{Q}}$ is \textit{finitely generated}
(with no \textit{syzygies}) by a unique primitive, homogeneous polynomial $%
\mathcal{I}\equiv \mathcal{I}\left( \mathcal{Q}\right) $, of degree two or
four in $\mathcal{Q}$ (which we will denotes as $\mathcal{I}_{2}$ resp. $%
\mathcal{I}_{4}$); see e.g. Table II of \cite{Kac-80}, and Refs. therein. In
all these cases, the formula (\ref{BH-entropy-area-formula}) acquires a
manifestly $G$-invariant form,%
\begin{equation}
S_{BH}=\pi \frac{A_{H}}{4}=\pi \left\{
\begin{array}{l}
\left\vert \mathcal{I}_{2}\left( \mathcal{Q}\right) \right\vert ; \\
\text{\textit{or}} \\
\sqrt{\left\vert \mathcal{I}_{4}\left( \mathcal{Q}\right) \right\vert }.%
\end{array}%
\right.   \label{BH-entropy-area-formula-E7}
\end{equation}%
Interestingly, formula (\ref{BH-entropy-area-formula-E7}) relates the
Bekenstein-Hawking entropy of extremal black holes to the theory of the
aforementioned distinguished class of $\theta $-groups, which can actually
be identified as Lie groups of type $E_{7}$ \textit{\`{a} la Brown} of
non-degenerate (when $\mathcal{I}=\mathcal{I}_{4}$) \cite{Brown} or
degenerate (when $\mathcal{I}=\mathcal{I}_{2}$) \cite{Garibaldi, FKM-deg-E7}
type. After Brown \cite{Brown}, non-degenerate groups of type $E_{7}$ can
always be characterized as automorphism groups of \textit{Freudenthal triple
systems} (which in turn can be of reduced or non-reduced type; see below).

Clearly, the value acquired by $\mathcal{I}$ is constant along any $G$%
-orbit. When $\mathcal{I}\neq 0$, the corresponding (generic, open,
non-nilpotent) $G$-orbit supports a \textquotedblleft
large\textquotedblright\ extremal black hole, which has $S_{BH}\neq 0$, and
thus $A_{H}\neq 0$, at the two-derivative (Einstein) level; on the other
hand, when $\mathcal{I}=0$, the corresponding (nilpotent) $G$-orbit supports
a \textquotedblleft small\textquotedblright\ extremal black hole, which has $%
S_{BH}=0$, and thus $A_{H}=0$, at the two-derivative (Einstein) level :
thus, such a \textquotedblleft small\textquotedblright\ black hole is
intrinsically quantum, since it needs of an higher-derivative theory of
gravity (such as the ones occurring in string effective actions) for a
sensible description as solution within a Lagrangian theory.

Moreover, in all the above cases, a manifestly $G$-invariant presentation of
the $G$-orbit stratification of $\mathbf{R}_{\mathcal{Q}}$ is given by the $%
1:1$ correpondence between $G$-invariant sets of algebro-differential
constraints on $\mathcal{I}\left( \mathcal{Q}\right) $ and the various
(classes of isomorphic) $G$-orbits $\mathcal{O}$'s. Over $\mathbb{C}$, all
\textquotedblleft large\textquotedblright\ $G$-orbits, which are level
hypersurfaces in $\mathbf{R}_{\mathcal{Q}}$ are isomorphic, defining the
generic, open orbit; however, over $\mathbb{R}$, different real forms (of
Riemannian or pseudo-Riemannian type) exist, distinguished by sign$\left(
\mathcal{I}\right) $, but possibly (when $G$ is non-degenerate and
non-split) also by further $G$-invariant \textquotedblleft
finer\textquotedblright\ constraints on $\mathcal{I}$. On the other hand,
when $G$ is non-degenerate, the stratification of \textquotedblleft small"
(i.e. nilpotent) $G$-orbits over $\mathbb{C}$ may involve $G$-invariant
differential constraints on $\mathcal{I}$, and, when $G$ is non-split, finer
splittings of the $G$-orbit stratification may occur over $\mathbb{R}$. For
instance, when $\mathcal{I}=\mathcal{I}_{4}$ (i.e., for $G$ being
non-degenerate of type $E_{7}$), the stratification of nilpotent $G$-orbits
is given by \cite{strat-pol}%
\begin{equation}
\begin{array}{lll}
\text{nilp.~}G\text{-orbit} & G\text{-inv.~constraint} & \text{rank}%
_{FTS}\left( \mathcal{Q}\right)  \\
\mathcal{O}_{3}: & \mathcal{I}_{4}=0 & 3 \\
\mathcal{O}_{2}: & \partial \mathcal{I}_{4}=0 & 2 \\
\mathcal{O}_{1}: & \left. \partial ^{2}\mathcal{I}_{4}\right\vert _{\mathbf{%
Adj}(G)}=0 & 1,%
\end{array}
\label{strat-C}
\end{equation}%
where rank$_{FTS}\left( \mathcal{Q}\right) $ indicates the $G$-invariant rank%
\footnote{%
If $\mathcal{Q}$ belongs to a \textquotedblleft large\textquotedblright\ $G$%
-orbit, i.e. when it is such that $\mathcal{I}_{4}(\mathcal{Q})\neq 0$, then
rank$_{FTS}(\mathcal{Q})=4.$} of $\mathcal{Q}\equiv \mathbf{R}_{\mathcal{Q}}$
as element of a (reduced) \textit{Freudenthal triple system} \cite%
{Krutelevich1,Krutelevich2}, which in turn is constructed over a rank-3 Jordan algebra
(which, for $\mathcal{N}=2$ \textit{symmetric} supergravities, are simple or
semi-simple; see table 2). Over $\mathbb{R}$, when $G$ is split, the
stratification of nilpotent orbits is still given by (\ref{strat-C}),
whereas when $G$ is minimally non-compact, each of the $\mathcal{O}_{3}$ and
$\mathcal{O}_{2}$ split into two $G$-orbits. Note that $\mathcal{O}_{1}$,
which is the minimal, highest weight $G$-orbit, has the \textit{largest}
stabilizer and it is always \textit{unique}.

\section{$\mathcal{N}=2$ \textit{symmetric} supergravities\label{Sect2}}

$\mathcal{N}=2$-extended Maxwell-Einstein supergravity theories \cite{GST1}-\nocite{GST2}\cite{GST3}
with homogeneous symmetric special K\"{a}hler vector multiplets' scalar
manifolds will henceforth be shortly referred to as \textit{symmetric}
Maxwell-Einstein supergravities. The Riemannian, non-compact, symmetric
special K\"{a}hler manifolds have the general coset structure%
\begin{equation}
M_{n_{V}}:=\frac{G}{H_{0}\otimes U(1)},  \label{coset}
\end{equation}%
where $H_{0}\otimes U\left( 1\right) $ is the maximal compact subgroup (mcs)
of the $U$-duality group $G$. They have been classified in \cite{CVP,dWVVP}
(see \textit{e.g.} \cite{LA08} for a quite recent account), and they are
reported in Table 1. All the corresponding supergravity theories actually
have a five-dimensional origin, since they can be obtained from
\textquotedblleft parent\textquotedblright\ (minimally supersymmetric) $%
\mathcal{N}=2$ supergravities in $4+1$ space-time dimensions, by
compactifying \textit{\`{a} la Kaluza-Klein} on $S^{1}$, and retaining the
massless sector. This is reflected in the fact that all such theories are
endowed with a holomorphic prepotential function which, after
projectivization of the coordinates, is a homogeneous cubic polynomial \cite{GST1}-\nocite{GST2}\cite{GST3}.
 The unique exception is provided by the so-called \textit{Luciani
theories} \cite{Luciani}, which do not have a five-dimensional origin and
correspond to the \textit{minimal coupling} of vector multiplets to $%
\mathcal{N}=2$ supergravity. The corresponding special K\"{a}hler manifolds
are all symmetric spaces, all with geodesic rank one, and they are nothing
but the Riemannian non-compact versions of the $n_{V}$-dimensional complex
projective spaces $\overline{\mathbb{C}P}^{n_{V}}$ (see e.g. \cite%
{BFGM1,Gnecchi-1}); in these theories, the prepotential is a homogeneous
quadratic polynomial, and thus the trilinear coupling of $\mathcal{N}=2$
supergravity, expressed by the so-called $C$-tensor of special geometry,
vanishes : $C_{ijk}=0$.

\begin{table}[h!]
\par
\begin{center}
\begin{tabular}{|c||c|c|c|}
\hline & $
\begin{array}{c}
\\
\frac{G}{H_{0}\times U(1)} \\
~
\end{array}
$ & $
\begin{array}{c}
\\
r \\
~
\end{array}
$ & $
\begin{array}{c}
\\
$dim$_{\mathbb{C}}\equiv n_{V} \\
~
\end{array}
$ \\ \hline\hline $
\begin{array}{c}
\textit{minimal~coupling} \\
n\in \mathbb{N}
\end{array}
$ & $\mathbb{CP}^{n}\equiv \frac{SU(1,n)}{U(1)\times SU(n)}~$ & $1$
& $n~$ \\ \hline $
\begin{array}{c}
\\
\mathbb{R}\oplus \mathbf{\Gamma }_{1,n-1},~n\in \mathbb{N} \\
~
\end{array}
$ & $\frac{SU(1,1)}{U(1)}\times \frac{SO(2,n)}{SO(2)\times SO(n)}~$
& $
\begin{array}{c}
\\
2~(n=1) \\
3~(n\geqslant 2)
\end{array}
~$ & $n+1$ \\ \hline $
\begin{array}{c}
\\
J_{3}^{\mathbb{O}} \\
~
\end{array}
$ & $\frac{E_{7(-25)}}{E_{6(-78)}\times U(1)}$ & $3$ & $27$ \\
\hline $
\begin{array}{c}
\\
J_{3}^{\mathbb{H}} \\
~
\end{array}
$ & $\frac{SO^{\ast }(12)}{U(6)}~$ & $3$ & $15$ \\ \hline $
\begin{array}{c}
\\
J_{3}^{\mathbb{C}} \\
~
\end{array}
$ & $\frac{SU(3,3)}{S\left( U(3)\times U(3)\right) }$ & $3$ & $9~$
\\ \hline $
\begin{array}{c}
\\
J_{3}^{\mathbb{R}} \\
~
\end{array}
$ & $\frac{Sp(6,\mathbb{R})}{U(3)}$ & $3$ & $6$
\\ \hline $
\begin{array}{c}
\\
\mathbb{R} \\
~
\end{array}
$ & $\frac{SL(2,\mathbb{R})}{U(1)}$ & $1$ & $1$
\\ \hline
\end{tabular}
\end{center}
\caption{\textbf{Riemannian symmetric} \textbf{non-compact
special K\"{a}hler spaces (\textit{alias} vector multiplets' scalar
manifolds of the \textit{symmetric} }$\mathcal{N}=2$\textbf{,}
$D=4$\textbf{\ Maxwell Einstein supergravity theories).
}$r$\textbf{\ denotes the geodesic rank of the manifold, whereas
}$n_{V}$\textbf{\ stands for the number of vector multiplets}}
\end{table}

As unraveled for the first time in \cite{GST1}-\nocite{GST2}\cite{GST3}, the cubic prepotentials of
\textit{symmetric} Maxwell-Einstein supergravities are all related to the
degree-3 (cubic) norm defined in the corresponding rank-3 Jordan algebra.
The sequence of factorized spaces in the third row of Table 1, which is
usually referred to as the \textit{generic Jordan family}, is related to the
semi-simple rank-3 Jordan algebras $\mathbb{R}\oplus \mathbf{\Gamma }_{1,n-1}
$, where $\mathbf{\Gamma }_{1,n-1}$ stands for the degree-$2$ Jordan algebra
with a quadratic form of Lorentzian signature $\left( 1,n-1\right) $ (%
\textit{spin factors}) \cite{Jordan}. The complex dimension of the
corresponding special K\"{a}hler manifold\footnote{%
This is the unique special K\"{a}hler manifold which is the product of two
irreducible spaces, as proved in \cite{Ferrara-VP}.}%
\begin{equation}
\frac{SL\left( 2,\mathbb{R}\right) }{U(1)}\otimes \frac{SO(2,n)}{%
SO(2)\otimes SO(n)}
\end{equation}%
is $n+1$, and its geodesic rank is $1+$min$\left( 2,n\right) $. On the other
hand, the four isolated \textit{\textquotedblleft magic\textquotedblright }
theories are based on the four simple rank-3 Jordan algebras $J_{3}^{\mathbb{%
O}}$, $J_{3}^{\mathbb{H}}$, $J_{3}^{\mathbb{C}}$ and $J_{3}^{\mathbb{R}}$,
which can be realized as generalized matrix algebras of $3\times 3$
Hermitian matrices over the four Huwitz's normed division algebras $\mathbb{O%
}$ (octonions), $\mathbb{H}$ (quaternions), $\mathbb{C}$ (complex numbers)
and $\mathbb{R}$ (real numbers) \cite{GST1,GST2,GST3,Jordan,Jacobson,Guna1,GPR}. The
name \textit{\textquotedblleft magic\textquotedblright } is due to the fact
that the Lie algebras of their $U$-duality groups in $D=2+1$, $3+1$ and $4+1$
space-time dimensions fit into the celebrated \textit{Magic Square} of
Freudenthal and Tits \cite{Freudenthal2,magic1,magic2}. By defining $A\equiv $dim$_{%
\mathbb{R}}\mathbb{A}$ ($=8,4,2,1$ for $\mathbb{A}=\mathbb{O},\mathbb{H},%
\mathbb{C},\mathbb{R}$, respectively), the complex dimension of the
symmetric cosets of the \textquotedblleft magic\textquotedblright\
supergravities is $3\left( A+1\right) $. \textit{Last but not least}, the
special K\"{a}hler scalar manifold of the so-called $T^{3}$-model is the
rank-1 coset $\frac{SL\left( 2,\mathbb{R}\right) }{U(1)}$ based on the cubic
prepotential $F=T^{3}$ and related to the simplest cubic Jordan algebra,
given by the real numbers, with the cubic norm simply given by the cube
power. This model has a unique vector multiplet coupled to $\mathcal{N}=2$
supergravity, and it can be obtained by dimensional reduction of
five-dimensional minimal \textquotedblleft pure\textquotedblright\
supergravity.

\subsection{\textquotedblleft Large\textquotedblright\ $U$-duality orbits}

The classification of $U$-duality orbits supporting \textquotedblleft
large\textquotedblright\ extremal black holes in symmetric Maxwell-Einstein
supergravities in $3+1$ space-time dimensions has been carried out in \cite%
{BFGM1}, and it is reported in\footnote{%
In the present report, we will not consider the highly-degenerate case given
by the so-called $T^{3}$-model, for which the reader is addressed to \cite%
{small orbits}, and to Refs. therein.} Table 2.

\begin{table}[h!]
\begin{center}
\begin{tabular}{|c||c|c|c|}
\hline & $
\begin{array}{c}
\\
\frac{1}{2}\text{-BPS orbit } \\
~~\mathcal{O}_{\frac{1}{2}-BPS}=\frac{G}{H_{0}} \\
~
\end{array}
$ & $
\begin{array}{c}
\\
\text{nBPS }Z_{H}\neq 0\text{ orbit} \\
\mathcal{O}_{nBPS,Z_{H}\neq 0}=\frac{G}{\widehat{H}}~ \\
~
\end{array}
$ & $
\begin{array}{c}
\\
\text{nBPS }Z_{H}=0\text{ orbit} \\
\mathcal{O}_{nBPS,Z_{H}=0}=\frac{G}{\widetilde{H}}~ \\
~
\end{array}
$ \\ \hline\hline $
\begin{array}{c}
\\
\textit{minimal~coupling} \\
~n\in \mathbb{N}
\end{array}
$ & $\frac{SU(1,n)}{SU(n)}~$ & $-$ & $\frac{SU(1,n)}{SU(1,n-1)}~$ \\
\hline $
\begin{array}{c}
\\
\mathbb{R}\oplus \mathbf{\Gamma }_{1,n-1} \\
~n\in \mathbb{N}
\end{array}
$ & $SU(1,1)\times \frac{SO(2,n)}{SO(2)\times SO(n)}~$ &
$SU(1,1)\times
\frac{SO(2,n)}{SO(1,1)\times SO(1,n-1)}~$ & $SU(1,1)\times \frac{SO(2,n)}{%
SO(2)\times SO(2,n-2)}$ \\ \hline $
\begin{array}{c}
\\
J_{3}^{\mathbb{O}} \\
~
\end{array}
$ & $\frac{E_{7(-25)}}{E_{6}}$ & $\frac{E_{7(-25)}}{E_{6(-26)}}$ & $\frac{%
E_{7(-25)}}{E_{6(-14)}}~$ \\ \hline $
\begin{array}{c}
\\
J_{3}^{\mathbb{H}} \\
~
\end{array}
$ & $\frac{SO^{\ast }(12)}{SU(6)}~$ & $\frac{SO^{\ast
}(12)}{SU^{\ast }(6)}~$ & $\frac{SO^{\ast }(12)}{SU(4,2)}~$ \\
\hline $
\begin{array}{c}
\\
J_{3}^{\mathbb{C}} \\
~
\end{array}
$ & $\frac{SU(3,3)}{SU(3)\times SU(3)}$ &
$\frac{SU(3,3)}{SL(3,\mathbb{C})}$ & $\frac{SU(3,3)}{SU(2,1)\times
SU(1,2)}~$ \\ \hline $
\begin{array}{c}
\\
J_{3}^{\mathbb{R}} \\
~
\end{array}
$ & $\frac{Sp(6,\mathbb{R})}{SU(3)}$ & $\frac{Sp(6,\mathbb{R})}{SL(3,\mathbb{%
R})}$ & $\frac{Sp(6,\mathbb{R})}{SU(2,1)}$ \\ \hline
\end{tabular}
\end{center}
\caption{\textbf{\textquotedblleft Large\textquotedblright\ $G$-orbits of \textit{symmetric }}$\mathcal{N}$\textbf{$=2$, $%
D=4$ Maxwell-Einstein supergravities. They all support extremal black hole attractors, with different supersymmetry-preserving features}}
\end{table}

Given the scalar manifold (\ref{coset}), the U-duality orbits which support (%
$\frac{1}{2}$-)BPS-saturated black holes , i.e. which preserve the maximal ($%
\frac{1}{2}$) amount of supersymmetry, has structure
\begin{equation}
\mathcal{O}_{BPS}=\frac{G}{H_{0}}\text{,~with~}H_{0}\otimes U(1)\overset{%
\text{mcs}}{\subsetneq }G.  \label{O_BPS-N=2}
\end{equation}%
As discovered in \cite{BFGM1}, there are other two non-isomorphic classes
of $U$-duality orbits, both supporting extremal black hole attractors which
are non-supersymmetric (i.e., which dor not saturate the BPS bound \cite{BPS}%
). The first non-supersymmetric (non-BPS) orbit has non-vanishing $\mathcal{N%
}=2$ central charge at the horizon ($Z_{H}\neq 0$), with coset structure
\begin{equation}
\mathcal{O}_{nBPS,Z_{H}\neq 0}=\frac{G}{\widehat{H}}\text{,~with~}\widehat{H}%
\otimes SO\left( 1,1\right) \subsetneq G,  \label{O_nBPS-Z<>0}
\end{equation}%
where $\widehat{H}$ denotes the $U$-duality group of the corresponding
parent theory in $4+1$ space-time dimensions, and $SO\left( 1,1\right) $
corresponds to the radius of the circle $S^{1}$ in the Kaluza-Klein
reduction from five to four dimensions. The second class of non-BPS $U$%
-duality orbits has vanishing central charge at the black hole horizon : $%
Z_{H}=0$, with coset structure
\begin{equation}
\mathcal{O}_{nBPS,Z_{H}=0}=\frac{G}{\widetilde{H}}\text{,~with~}\widetilde{H}%
\otimes U\left( 1\right) \subsetneq G.  \label{O_nBPS-Z=0}
\end{equation}%
Note that $\widehat{H}$ and $\widetilde{H}$ are the only two non-compact
forms of $H_{0}$ embedded (with rank-1 commutant) into $G$ itself. Thus, the
group embedding in the r.h.s. of (\ref{O_nBPS-Z<>0}) and (\ref{O_nBPS-Z=0})
are both maximal and symmetric (see \textit{e.g.} \cite%
{Gilmore,Helgason,Slansky}).

While $H_{0}$ is a real compact Lie group (stabilizing the BPS
\textquotedblleft large\textquotedblright\ orbit (\ref{O_BPS-N=2})), the
groups $\widehat{H}$ and $\widetilde{H}$, respectively stabilizing the
non-BPS \textquotedblleft large\textquotedblright\ orbits with $Z_{H}\neq 0$
(\ref{O_nBPS-Z<>0}) and $Z_{H}=0$ (\ref{O_nBPS-Z=0}), are non-compact, and
thus they will admit a proper maximal compact subgroup, which we denote with
$\widehat{h}$ resp. $\widetilde{h}$ :
\begin{equation}
\widehat{h}=\text{mcs}\left( \widehat{H}\right) ;~~\widetilde{h}=\text{mcs}%
\left( \widetilde{H}\right) .  \label{N=2-mcs's}
\end{equation}

\subsection{\textquotedblleft Moduli spaces\textquotedblright\ of attractors}

For symmetric $\mathcal{N}=2$ supergravities, general results on the rank $%
\mathfrak{r}$ of the $2n_{V}\times 2n_{V}$ Hessian matrix $\mathbf{H}$ of
the effective black hole potential $V_{BH}\left( \phi ,\mathcal{Q}\right) $
at its critical points are known (see e.g. \cite{BFGM1} and \cite{K-rev}).

The BPS (non-degenerate) critical points of $V_{BH}$ are stable, and thus
the Hessian matrix at BPS critical points $\mathbf{H}_{BPS}$ has no massless
modes \cite{FGK}, and its rank is maximal: $\mathfrak{r}_{BPS}=2n_{V}$.
Furthermore, the analysis carried out in \cite{BFGM1} showed that for the
other two classes of non-BPS critical points of $V_{BH}$, the rank of $%
\mathbf{H}$ is model-dependent:
\begin{eqnarray}
\mathbb{CP}^{n} &:&\mathfrak{r}_{nBPS,Z_{H}=0}=2; \\
\mathbb{R}\oplus \mathbf{\Gamma }_{1,n-1} &:&\left\{
\begin{array}{l}
\mathfrak{r}_{nBPS,Z_{H}\neq 0}=n+2; \\
\\
\mathfrak{r}_{nBPS,Z_{H}=0}=6;%
\end{array}%
\right.  \\
J_{3}^{\mathbb{A}} &:&\left\{
\begin{array}{l}
\mathfrak{r}_{nBPS,Z_{H}\neq 0}=3A+4; \\
\\
\mathfrak{r}_{nBPS,Z_{H}=0}=2A+6.%
\end{array}%
\right.
\end{eqnarray}%
Correspondingly, the number $\sharp $ of massless Hessian modes for the
various models is given by%
\begin{equation}
\sharp :=2n_{V}-\mathfrak{r},
\end{equation}%
and thus%
\begin{eqnarray}
\mathbb{CP}^{n} &:&\sharp _{nBPS,Z_{H}=0}=2\left( n_{V}-1\right) ; \\
\mathbb{R}\oplus \mathbf{\Gamma }_{1,n-1} &:&\left\{
\begin{array}{l}
\sharp _{nBPS,Z_{H}\neq 0}=n; \\
\\
\sharp _{nBPS,Z_{H}=0}=2n-4;%
\end{array}%
\right.  \\
J_{3}^{\mathbb{A}} &:&\left\{
\begin{array}{l}
\sharp _{nBPS,Z_{H}\neq 0}=3A+2; \\
\\
\sharp _{nBPS,Z_{H}=0}=4A.%
\end{array}%
\right.
\end{eqnarray}%
From previous statements, it also holds that%
\begin{equation}
\sharp _{BPS}=0\label{BPS}
\end{equation}%
for all $\mathcal{N}=2$ theories, regardless the properties of the special K%
\"{a}hler vector multiplets' scalar manifold.\medskip

Let us start by recalling that $V_{BH}$ is defined as
\begin{equation}
V_{BH}\left( \phi ,\mathcal{Q}\right) :=-\frac{1}{2}\mathcal{Q}^{T}\mathbf{M}%
(\phi )\mathcal{Q},  \label{V_BH}
\end{equation}%
where $\phi $ denotes the $2n_{V}$ real scalar fields parametrising the
special K\"{a}hler scalar manifold $\frac{G}{H_{0}\otimes U(1)}$, and $%
\mathcal{Q}$ is the symplectic vector of e.m. black hole charges sitting in
the $G$-irrepr. $\mathbf{R}_{\mathcal{Q}}$ of the $U$-duality group $G$.
Moreover, $\mathbf{M}(\phi )$ is the $2\left( n_{V}+1\right) \times 2\left(
n_{V}+1\right) $ real, symmetric and symplectic matrix defined as \cite%
{33,34,d-geom-rev}
\begin{equation}
\mathbf{M}(\phi )=-\left( \mathbf{LL}^{T}\right) ^{-1},
\end{equation}%
where $\mathbf{L}=\mathbf{L}\left( \phi \right) $ is coset representative of
$\frac{G}{H_{0}\otimes U(1)}$, \textit{i.e.} a local section of the
principal $G$-bundle over the special Hodge-K\"{a}hler scalar manifold $%
\frac{G}{H_{0}\otimes U(1)}$, with structure group $H_{0}\otimes U(1)$.

The action of an element $g\in G$ on $V_{BH}$ (\ref{V_BH}) is such that
\begin{equation}
G:V_{BH}\left( \phi ,\mathcal{Q}\right) \mapsto V_{BH}\left( \phi _{g},%
\mathcal{Q}^{g}\right) =V_{BH}\left( \phi _{g},\left( g^{-1}\right) ^{T}%
\mathcal{Q}\right) ;
\end{equation}%
thus, $V_{BH}$ is not $G$-invariant, because its coefficients (given by the
components of $\mathcal{Q}$) do not in general remain the same. The
situation changes if one restricts $g$ to $g_{\mathcal{Q}}\in \mathcal{H}$,
i.e. if one restricts to the stabilizer $\mathcal{H}$ of the
\textquotedblleft large\textquotedblright\ $G$-orbits $\mathcal{O}\simeq
\frac{G}{\mathcal{H}}\subsetneq \mathbf{R}_{\mathcal{Q}}$ (cf. (\ref{orb}))
to which $\mathcal{Q}$ belongs. In such a case, by definition of $\mathcal{H}
$ :
\begin{gather}
\mathcal{Q}^{g_{\mathcal{Q}}}=\mathcal{Q}; \\
\Downarrow  \notag \\
\mathcal{H}:V_{BH}\left( \phi ,\mathcal{Q}\right) \mapsto V_{BH}\left( \phi
_{g_{\mathcal{Q}}},\mathcal{Q}^{g_{\mathcal{Q}}}\right) =V_{BH}\left( \phi
_{g_{\mathcal{Q}}},\left( g_{\mathcal{Q}}^{-1}\right) ^{T}\mathcal{Q}\right)
=V_{BH}\left( \phi _{g_{\mathcal{Q}}},\mathcal{Q}\right) \simeq V_{BH}\left(
\phi ,\mathcal{Q}\right) .  \label{tthis}
\end{gather}

Then, it is natural to split the $2n_{V}$ real scalar fields $\phi $ as $%
\phi =\left\{ \phi _{\mathcal{Q}},\breve{\phi}_{\mathcal{Q}}\right\} $, where

\begin{itemize}
\item
\begin{equation}
\phi _{\mathcal{Q}}\in \frac{\mathcal{H}}{\text{mcs}\left( \mathcal{H}%
\right) }\subsetneq M_{n_{V}},
\end{equation}%
where we denote%
\begin{equation}
\frac{\mathcal{H}}{\text{mcs}\left( \mathcal{H}\right) }=:\mathcal{M}_{%
\mathcal{Q}};\label{mod}
\end{equation}

\item $\breve{\phi}_{\mathcal{Q}}$ coordinatize the complement of $\mathcal{M%
}_{\mathcal{Q}}$ in $M_{n_{V}}$ :%
\begin{equation}
\breve{\phi}_{\mathcal{Q}}\in M_{n_{V}}\backslash \mathcal{M}_{\mathcal{Q}}.
\end{equation}
\end{itemize}

One can then define%
\begin{equation}
V_{BH,crit}\left( \phi _{\mathcal{Q}},\mathcal{Q}\right) :=\left.
V_{BH}\left( \phi ,\mathcal{Q}\right) \right\vert _{\frac{\partial V_{BH}}{%
\partial \breve{\phi}_{\mathcal{Q}}}=0}\left( \neq 0\right)
\end{equation}%
as the values of $V_{BH}$ along the equations of motion for the scalars $%
\breve{\phi}_{\mathcal{Q}}$. Thus, (\ref{tthis}) implies the invariance of $%
V_{BH,crit}\left( \phi _{\mathcal{Q}},\mathcal{Q}\right) $ under $\mathcal{H}
$ :
\begin{equation}
\mathcal{H}:V_{BH,crit}\left( \phi _{\mathcal{Q}},\mathcal{Q}\right) \mapsto
V_{BH,crit}\left( \left( \phi _{\mathcal{Q}}\right) _{g_{\mathcal{Q}}},%
\mathcal{Q}\right) \simeq V_{BH,crit}\left( \phi _{\mathcal{Q}},\mathcal{Q}%
\right)
\end{equation}

Finally, it is crucial to observe that $\mathcal{H}$, except for the ($\frac{%
1}{2}$-)BPS \textquotedblleft large\textquotedblright\ $G$-orbit, is
generally a \textit{non-compact} real Lie group. This implies that $V_{BH}$
at its critical points is independent on the subset of scalar fields
\begin{equation}
\phi _{\mathcal{Q}}\in \mathcal{M}_{\mathcal{Q}}\subsetneq M_{n_{V}},
\end{equation}%
i.e. on those scalar fields belonging to the homogeneous symmetric
submanifold $\mathcal{M}_{\mathcal{Q}}\subsetneq M_{n_{V}}$, which thus be
regarded as the \textquotedblleft moduli space\textquotedblright\ of the
attractor solutions supported by the charge orbit $\mathcal{O}\simeq \frac{G%
}{\mathcal{H}}\subsetneq \mathbf{R}_{\mathcal{Q}}$. Thus,%
\begin{equation}
\frac{\partial V_{BH,crit}\left( \phi _{\mathcal{Q}},\mathcal{Q}\right) }{%
\partial \phi _{\mathcal{Q}}}=0\Rightarrow \left. V_{BH}\left( \phi ,%
\mathcal{Q}\right) \right\vert _{\frac{\partial V_{BH}}{\partial \breve{\phi}%
_{\mathcal{Q}}}=0}=:V_{BH,crit}\left( \mathcal{Q}\right) ,
\end{equation}%
or, equivalently :%
\begin{equation}
\left. V_{BH}\left( \phi ,\mathcal{Q}\right) \right\vert _{\frac{\partial
V_{BH}}{\partial \phi }=0}=\left. V_{BH}\left( \phi ,\mathcal{Q}\right)
\right\vert _{\frac{\partial V_{BH}}{\partial \breve{\phi}_{\mathcal{Q}}}%
=0}=:V_{BH,crit}\left( \mathcal{Q}\right) .
\end{equation}%
By using this line of reasoning, in \cite{Ferrara-Marrani-2} (see also \cite%
{Ferrara-Marrani-1}) it was proved that, remarkably, the rank of $\mathbf{H}$
corresponds to all positive eigenvalues (i.e., stable directions in the
scalar manifold), and also that the massless modes of $\mathbf{H}$ are
actually \textquotedblleft flat\textquotedblright\ directions of $V_{BH}$ at
the corresponding classes of its critical points. Thus, such
\textquotedblleft flat\textquotedblright\ directions of the critical values
of $V_{BH}$ span some \textquotedblleft moduli spaces\textquotedblright\ of
the attractor solutions \cite{Ferrara-Marrani-2}, corresponding to those
scalar degrees of freedom which are not stabilized by the AM at the horizon
of the extremal black hole. The general coset structure of such
\textquotedblleft moduli spaces\textquotedblright\ has the orbit stabilizer
as global isometry, and its corresponding mcs as isotropy group; thus, by
virtue of the treatment above (cf. (\ref{mod})), one can generally write that%
\begin{eqnarray}
\mathcal{M}_{BPS} &=&\frac{H_{0}}{\text{mcs}\left( H_{0}\right) }\simeq
\varnothing ;  \label{uno} \\
\mathcal{M}_{nBPS,Z\neq 0} &=&\frac{\widehat{H}}{\text{mcs}\left( \widehat{H}%
\right) }=\frac{\widehat{H}}{\widehat{h}},~~\text{dim}_{\mathbb{R}}\left(
\mathcal{M}_{nBPS,Z\neq 0}\right) =\sharp _{nBPS,Z\neq 0};  \label{due} \\
\mathcal{M}_{nBPS,Z=0} &=&\frac{\widetilde{H}}{\text{mcs}\left( \widetilde{H}%
\right) }=\frac{\widetilde{H}}{\widetilde{h}},~~\text{dim}_{\mathbb{R}%
}\left( \mathcal{M}_{nBPS,Z=0}\right) =\sharp _{nBPS,Z=0},  \label{tre}
\end{eqnarray}%
where the non-existence of $\mathcal{M}_{BPS}$ follows from (\ref{BPS}).
This means that in $\mathcal{N}=2$ symmetric supergravities all critical
points of $V_{BH}$ supported by \textquotedblleft large\textquotedblright\ $U
$-duality orbits are \textit{stable}, up to a (possibly vanishing) certain
number $\sharp $ of \textquotedblleft flat\textquotedblright\ directions,
spanning some proper subspace of the scalar manifold itself :%
\begin{eqnarray}
\mathcal{M}_{nBPS,Z\neq 0} &\subsetneq &M_{n_{V}}; \\
\mathcal{M}_{nBPS,Z=0} &\subsetneq &M_{n_{V}}.
\end{eqnarray}%
Tables 3 and 4 report spaces $\mathcal{M}_{nBPS,Z\neq 0}$ and $\mathcal{M}%
_{nBPS,Z=0}$, respectively \cite{Ferrara-Marrani-2}.

\begin{table}[h!]
\par
\begin{center}
\begin{tabular}{|c||c|c|c|}
\hline & $
\begin{array}{c}
\\
\frac{\widehat{H}}{mcs(\widehat{H})} \\
~
\end{array}
$ & $
\begin{array}{c}
\\
r \\
~
\end{array}
$ & $
\begin{array}{c}
\\
$dim$_{\mathbb{R}} \\
~
\end{array}
$ \\ \hline $
\begin{array}{c}
\\
\mathbb{R}\oplus \mathbf{\Gamma }_{1,n-1},~n\in \mathbb{N} \\
~
\end{array}
$ & $SO(1,1)\times \frac{SO(1,n-1)}{SO(n-1)}~$ & $
\begin{array}{c}
\\
1~(n=1) \\
2~(n\geqslant 2)
\end{array}
~$ & $n$ \\ \hline $
\begin{array}{c}
\\
J_{3}^{\mathbb{O}} \\
~
\end{array}
$ & $\frac{E_{6(-26)}}{F_{4(-52)}}$ & $2$ & $26$ \\ \hline $
\begin{array}{c}
\\
J_{3}^{\mathbb{H}} \\
~
\end{array}
$ & $\frac{SU^{\ast }(6)}{USp(6)}~$ & $2$ & $14$ \\ \hline $
\begin{array}{c}
\\
J_{3}^{\mathbb{C}} \\
~
\end{array}
$ & $\frac{SL(3,\mathbb{C})}{SU(3)}$ & $2$ & $8~$ \\ \hline $
\begin{array}{c}
\\
J_{3}^{\mathbb{R}} \\
~
\end{array}
$ & $\frac{SL(3,\mathbb{R})}{SO(3)}$ & $2$ & $5$ \\ \hline
\end{tabular}
\end{center}
\caption{``\textbf{Moduli spaces'' of non-BPS }$Z_{H}\neq0$ \textbf{extremal black hole attractors} \textbf{in }$\mathcal{N}$\textbf{$%
=2$, $D=4$ \textit{symmetric} Maxwell-Einstein supergravities. They are the real special (vector multiplets') scalar manifolds of the corresponding }%
$\mathcal{N}$$=2$\textbf{,} $D=5$ \textbf{ symmetric \textquotedblleft parent\textquotedblright\ supergravity theory}}
\end{table}

\begin{table}[h!]
\par
\begin{center}
\begin{tabular}{|c||c|c|c|}
\hline & $
\begin{array}{c}
\\
\frac{\widetilde{H}}{mcs(\widetilde{H})}\equiv\frac{\widetilde{H}}{\widetilde{h}%
^{\prime }\times U(1)} \\
~
\end{array}
$ & $
\begin{array}{c}
\\
r \\
~
\end{array}
$ & $
\begin{array}{c}
\\
$dim$_{\mathbb{C}} \\
~
\end{array}
$ \\ \hline\hline $
\begin{array}{c}
\text{\textit{minimal~coupling}} \\
n\in \mathbb{N}
\end{array}
$ & $\frac{SU(1,n-1)}{U(1)\times SU(n-1)}~$ & $1$ & $n~-1$ \\ \hline
$
\begin{array}{c}
\\
\mathbb{R}\oplus \mathbf{\Gamma }_{1,n-1},~n\in \mathbb{N} \\
~
\end{array}
$ & $\frac{SO(2,n-2)}{SO(2)\times SO(n-2)},~n\geqslant 3~$ & $
\begin{array}{c}
\\
1~(n=3) \\
2~(n\geqslant 4)
\end{array}
~$ & $n-2$ \\ \hline $
\begin{array}{c}
\\
J_{3}^{\mathbb{O}} \\
~
\end{array}
$ & $\frac{E_{6(-14)}}{SO(10)\times U(1)}$ & $2$ & $16$ \\ \hline $
\begin{array}{c}
\\
J_{3}^{\mathbb{H}} \\
~
\end{array}
$ & $\frac{SU(4,2)}{SU(4)\times SU(2)\times U(1)}~$ & $2$ & $8$ \\
\hline $
\begin{array}{c}
\\
J_{3}^{\mathbb{C}} \\
~
\end{array}
$ & $\frac{SU(2,1)}{SU(2)\times U(1)}\times
\frac{SU(1,2)}{SU(2)\times U(1)}$ & $2$ & $4$ \\ \hline $
\begin{array}{c}
\\
J_{3}^{\mathbb{R}} \\
~
\end{array}
$ & $\frac{SU(2,1)}{SU(2)\times U(1)}$ & $1$ & $2$ \\ \hline
\end{tabular}
\end{center}
\caption{``\textbf{Moduli spaces'' of non-BPS }$Z_{H}=0$\textbf{ extremal black hole attractors } \textbf{in }$\mathcal{N}$\textbf{$=2$, $%
D=4 $ \textit{symmetric} Maxwell-Einstein supergravities. They are
(non-special) symmetric K\"{a}hler manifolds}}
\end{table}

Interestingly, the \textquotedblleft moduli space\textquotedblright\ $%
\mathcal{M}_{nBPS,Z\neq 0} $ of non-BPS $Z_{H}\neq 0$ attractors is the
scalar manifold of the corresponding \textquotedblleft
parent\textquotedblright\ theory in $4+1$ space-time dimensions \cite%
{Ferrara-Marrani-2} (see also \cite{CFM1} and \cite{TT2} for a result
holding for generic special $d$-geometries).

\subsection{\textquotedblleft Moduli spaces\textquotedblright\ of the ADM
mass}

Remarkably, by the thumb rule of orbit stabilizer modded by its mcs, one can
associate \textquotedblleft moduli spaces\textquotedblright\ also to
\textquotedblleft small\textquotedblright\ $U$-duality\ orbits, which do not
attractor black holes : indeed, as mentioned above, the corresponding black
hole has vanishing entropy in the Einstein, two-derivative apporximation,
and no AM (at least in the sense pointed out in the previous section; see
\cite{TVvR}) holds \cite{CFMZ1,Duff-N=8,CFMZ1-d=5,small orbits}. For
\textquotedblleft small\textquotedblright\ U-duality orbits there exists a
\textquotedblleft moduli space\textquotedblright\ also when the semi-simple
part of the orbit stabilizer is a compact real Lie group : in such cases,
the \textquotedblleft moduli space\textquotedblright\ is spanned by the
non-reductive, translational part of the orbit stabilizer itself \cite%
{CFMZ1-d=5,small orbits}. \textit{\c{C}a va sans dire} that for
\textquotedblleft small\textquotedblright\ orbits,\ there is no event
horizon of the extremal black hole at which the N=2 central charge should be
evaluated and no AM holds : in these cases, one may consider the
asymptotical, spacial limit of the black hole, and put forward the
interpretation of the \textquotedblleft moduli spaces\textquotedblright\
associated to \textquotedblleft small\textquotedblright\ orbits as
\textquotedblleft moduli spaces\textquotedblright of the ADM mass \cite{ADM}
of the \textquotedblleft small\textquotedblright\ black hole itself.

\end{document}